\documentstyle[preprint,aps]{revtex}
\tighten

\newcommand{\bc}{\begin{center}}
\newcommand{\ec}{\end{center}}
\newcommand{\beqn}{\begin{equation}}
\newcommand{\eeqn}{\end{equation}}
\newcommand{\barr}{\begin{eqnarray}}
\newcommand{\earr}{\end{eqnarray}}
\begin{document}
\begin{flushright}
June 1996
\end{flushright}
\vskip 2cm
\bc
\Large\bf Role of Dual Higgs Mechanism \\
in Chiral Phase Transition at Finite Temperature\\
\vskip 1cm
\ec
\centerline{\large Shoichi Sasaki, Hideo Suganuma and Hiroshi Toki}
\vskip 0.5cm

\centerline
{\large {\it Research Center for Nuclear Physics } {\rm (}\it RCNP {\rm )} }
\vskip 0.2cm
\centerline{\large \it Osaka University, Ibaraki, Osaka 567, Japan}
%%%%%%%%%%%%%%%%%%%  Abstract  %%%%%%%%%%%%%%%%%%%%%%%%%%%%%%%%%%%
\baselineskip 24pt
\vskip 1cm
\begin{abstract}
\baselineskip 20pt
\indent

The chiral phase transition at finite temperature is studied 
by using the Schwinger-Dyson equation
in the dual Ginzburg-Landau theory, in which the dual Higgs mechanism plays 
an essential role on both the color confinement and the spontaneous 
chiral-symmetry breaking. 
At zero temperature, quark condensate is strongly 
correlated with the string tension, 
which is generated by QCD-monopole condensation,
as $\langle {\bar q}q \rangle^{1/3} 
\stackrel{\propto}{\scriptstyle \sim} \sqrt{\sigma}$. 
In order to solve the finite-temperature 
Schwinger-Dyson equation numerically,
we provide a new ansatz for the quark self-energy 
in the imaginary-time formalism.
The recovery of the chiral symmetry is found at high temperature;
$T_{_{C}}\sim 100{\rm MeV}$ with realistic parameters.
We find also strong correlation between the critical temperature $T_{_{C}}$
of the chiral symmetry restoration and the strength of the string tension.

\end{abstract}
\newpage
%%%%%%%%%%%%%%%%%%%%%%%%%%%%%%%%%%%%%%%%%%%%%%%%%%%%%%%%%%%%%%%%%

\indent
In ${\rm SU}(N_{c})$ gauge theory, the appearance of magnetic monopoles was 
pointed out with the idea of the abelian gauge fixing based on the 
topological argument, 
$\pi_{2}({\rm SU}(N_{c})/{\rm U}(1)^{N_{c}-1})=Z^{N_{c}-1}_{\infty}$, by 't Hooft 
\cite{tHooft}. He conjectured that the dual Meissner effect 
would be realized if QCD-monopoles were condensed\cite{tHooft}.
The basic idea for color confinement is that
the color-electric flux between quarks is squeezed like a string or a tube, 
because the color-electric fields are excluded in the QCD-vacuum.
This is similar to the Meissner effect in the superconductor, where the 
ordinary magnetic fields are excluded.
The linear-confining potential is produced, since the squeezed flux 
has a uniform energy per unit length.
The recent lattice QCD simulations support this idea that QCD-monopoles play a 
crucial role on color confinement through their condensation 
\cite{Monopole}.

The effective theory of non-perturbative QCD may be derived from
QCD by using the abelian gauge fixing, 
which brings the non-abelian gauge theory into the 
abelian gauge theory with QCD-monopoles\cite{tHooft}.
In order to describe the above mentioned dual Meissner effect,
the dual Ginzburg-Landau (DGL) theory is constructed following 
Nambu's demonstration\cite{Nambu}, which uses the relativistic version of the 
Ginzburg-Landau theory on the Abrikosov vortex solution in the 
superconductor\cite{NO}.
QCD-monopole condensation causes strong and long range correlations 
between a quark and antiquark pair, which produce the linear-confining potential 
through the dual Higgs mechanism\cite{{kanazawa},{SST}}.
We assume in addition the abelian dominance, in which
the non-abelian part does not contribute to non-perturbative
phenomena at low energy and is neglected\cite{EI}.
We then studied chiral symmetry breaking in the DGL
theory\cite{{SST},{TSS},{COMO},{SST2}}.
It was found that QCD-monopole condensation plays an essential role 
on the dynamical mass generation of the quark, which means 
spontaneous chiral-symmetry breaking, 
by using the Schwinger-Dyson (SD) equation for the dynamical 
quark\cite{{SST},{TSS},{COMO},{SST2}}. 
We mention here that the lattice QCD study supports this idea of 
providing spontaneous chiral-symmetry breaking due to 
QCD-monopole condensation\cite{{Miya},{WO}}.

The high temperature and/or high density region would be realized in 
the laboratories as the intermediate states of 
the relativistic heavy-ion collisions, the study of which will be made 
at RHIC in BNL and at LHC in CERN in the near future.
The possible signals of a new phase, in which the liberation of colors, 
namely quarks and gluons, takes place, are expected to be seen
in these experiments.
Anticipating these experiments, the quark-gluon plasma (QGP) 
have been studied by many theorists.
It is necessary for both confinement-deconfinement phase transition and 
chiral phase transition to understand the QCD phase transition.
However, we do not have yet an effective theory
to study the QGP physics based on the systematic treatment of 
these phase transitions, 
except for the lattice QCD simulations,
which are not accessible for all the necessary physical quantities
at the present.

We study the QGP physics on the basis of the DGL
theory, which describes both color confinement and chiral symmetry 
breaking at zero temperature\cite{{SST},{TSS},{COMO},{SST2}}.
In particular, we concentrate on the manifestation of chiral symmetry at 
finite temperature with confinement properties, which are based on 
the dual Meissner effect in the QCD vacuum. 
In this respect, we would like to mention the successful use of the 
DGL theory on the confinement-deconfinement phase transition at finite 
temperature\cite{Ichie}.
We shall formulate the finite-temperature SD equation using the 
imaginary-time formalism.

The DGL theory is an infrared effective theory of 
non-perturbative QCD based on the dual Higgs mechanism in the abelian 
gauge\cite{{kanazawa},{SST}},
%
%  eq.(1)
%
\beqn
{\cal L}_{\rm DGL}
={\rm tr} {\hat K}_{\rm gauge}(A_{\mu},B_{\mu})
+\bar q(i\partial \kern -2mm / -e A \kern -2.5mm / -m_q)q 
+ {\rm tr} [{\hat {\bf \cal D}}_{\mu},\chi ]^{\dag} [{\hat {\bf \cal D}}^{\mu},\chi ]
-\lambda {\rm tr} (\chi^{\dag} \chi - v^{2})^{2}\;\;,
\eeqn
where ${\hat {\bf \cal D}}_{\mu} \equiv {\hat \partial}_{\mu} + i g 
B_{\mu}$ is the
dual covariant derivative. The dual gauge coupling $g$ obeys the Dirac
condition $eg = 4\pi$ with $e$ being the gauge coupling.
The diagonal gluon $A_{\mu}$ and the dual gauge 
field $B_{\mu}$ are defined on the Cartan subalgebra $T_{3}$, $T_{8}$ 
on $\rm SU(3)$: $A^{\mu}\equiv A_{3}^{\mu}T_{3}+A_{8}^{\mu}T_{8}$,
$B^{\mu}\equiv B_{3}^{\mu}T_{3}+B_{8}^{\mu}T_{8}$.
The QCD-monopole field $\chi$ is defined on the $\rm SU(3)$ root 
vectors $E_{\alpha}$: $\chi \equiv \sqrt{2} 
\sum^{3}_{\alpha=1}\chi_{\alpha}E_{\alpha}$.
${\hat K}_{\rm gauge}$ is the kinetic term of the gauge fields 
$(A_{\mu},B_{\mu})$ in the Zwanziger form\cite{Zwan},
%
%  eq.(2)
%
\beqn
{\hat K}_{\rm gauge} \equiv -{1 \over {n^{2}}}
[n\cdot(\partial \wedge A)]^{2}
-{1 \over {n^{2}}}[n\cdot(\partial \wedge B)]^{2}
-{2 \over {n^{2}}}[n\cdot(\partial \wedge A)]_{\nu}
[n\cdot^{\ast}(\partial \wedge B)]^{\nu}\;\;,
\eeqn
where the duality of the gauge theory is manifest. 
The DGL lagrangian has the $[{\rm U}(1)_{e}]^{2}\times[{\rm U}(1)_{m}]^{2}$
gauge symmetry.

It is straightforward to calculate the static confining potential 
using the static quark-antiquark pair sources, which are totally color-singlet, 
located at a distance $r=|{\bf b}-{\bf a}|$.
%
%  eq.(3)
%
\beqn
{\bf j}_{\mu}={\bf Q}g_{\mu 0}\{ \delta^{3}({\bf x} - {\bf b})
 - \delta^{3}({\bf x} - {\bf a}) \}\;\;,
\label{source}
\eeqn
where ${\bf Q}=( Q_{3}, Q_{8} )$ is the color charge of the static quark.  
The linear potential; $\sigma r$ in the long distance 
is derived from the quark color-electric 
current-current correlation; $-{1 \over 2}{\bf j}_{\mu}D^{\mu 
\nu}{\bf j}_{\nu}$ with the static source (\ref{source}).
Here, $D_{\mu \nu}$ denotes the diagonal gluon propagator.
We then get the string tension as the simple 
expression\cite{CONF95},
%
%  eq.(4)
%
\beqn
\sigma \simeq {{ {{\bf Q}^2 m_{B}^2} \over 8\pi} \over
{\sqrt{1 - {{3g^{2}} \over \lambda}}}}
\ln({{1+\sqrt{1 - {{3g^{2}} \over \lambda}}} \over 
{1-\sqrt{1 - {{3g^{2}} \over \lambda}}}})={{2\pi v^{2}\kappa}
\over
{\sqrt{\kappa^{2} - 2 }}}
\ln({{\kappa+\sqrt{\kappa^{2} - 2 }} \over 
{\kappa-\sqrt{\kappa^{2} - 2 }}})
\eeqn 
% %
with ${\bf Q}^{2}={\footnotesize {{N_c -1} \over {2 N_c}}}\cdot e^2
={\footnotesize {e^{2} \over 3}}$ for $N_{c}=3$
and $\kappa={\footnotesize \sqrt{{2\lambda} \over {3g^{2}}}}$ 
corresponding to the Ginzburg-Landau parameter.
The mass of $B_{\mu}$, $m_{B}=\sqrt{3}gv$, is proportional to the 
QCD-monopole condensate $v$\cite{{kanazawa},{SST}}.
For the type-II limit ($\kappa \gg \sqrt{2}$), one finds $\sigma \simeq 
2\pi v^{2} \ln(2\kappa^{2})$, corresponding to the
formula for the energy per unit length of the Abrikosov 
vortex in the type-II superconductor \cite{{SST},{TSS}}.
We may set parameters as $e=5.5$,
$\lambda = 25$ and $v=0.126{\rm GeV}$, 
which reproduce the string tension ($\sqrt{\sigma} \simeq 0.44 {\rm GeV}$) and the 
cylindrical radius of the hadron flux tube ($R\sim m_B^{-1} \simeq 0.4 {\rm fm}$)
\cite{SST}.

We get the ordinary SD equation by using the
non-perturbative gluon propagator, obtained through QCD-monopole condensation
in the DGL theory, as the full gluon propagator 
of QCD in order to include the non-perturbative effect in the infrared region.
Hereafter, we concentrate on the case of the chiral limit for the simplicity 
of the argument. The SD equation for the dynamical quark 
propagator $S_q(p)$ in the rainbow approximation is written as
%
%  eq.(5)
%
\beqn
S_q^{-1}(p)= i{p \kern -2mm /} + 
\int {{d^4k} \over (2\pi)^4} 
{\bf Q}^2 \gamma_\mu S_q(k) \gamma_\nu D_{\mu \nu}
(p-k) \;\;,
\label{sde}
\eeqn
in the Euclidean metric. 

In the QCD-monopole condensed vacuum, the non-perturbative gluon 
propagator is derived by integrating out 
$B_{\mu}$\cite{{SST},{TSS},{COMO},{SST2}},
%
%  eq.(6)
%
\beqn
D_{\mu \nu}(k)=-{1 \over k^2} \left( {\delta_{\mu \nu}+(\alpha_e -1)
{{k_\mu k_\nu} \over k^2}} \right) - {1 \over k^2}{m^2_B \over {k^2 + m^2_B}} \cdot
{{\epsilon_{\lambda \mu \alpha \beta} {\epsilon_{\lambda \nu \gamma \delta}}
n_\alpha n_\gamma k_\beta k_\delta} \over {{(n \cdot k)}^2 + a^2}} \;,
\label{GP}
\eeqn
where $\alpha_e$ is the gauge fixing parameter on the residual 
abelian gauge symmetry $[{\rm U}(1)_{e}]^{2}$. 
It is noted that this modified gluon propagator is asymptotically reduced to the 
original gluon propagator of QCD in the ultraviolet region; $k^{2} \gg m_{B}^{2}$.
We have introduced the infrared cutoff $a$ in this propagator (\ref{GP}) corresponding 
to the dynamical quark-antiquark pair creation and/or the color 
confinement in the size of 
hadrons\cite{SST2}.

We take the angular average on the direction of 
the Dirac string $n_\mu$ in the SD equation,
%
%  eq.(7)
%
\beqn
\left\langle {1 \over (n\cdot k)^2+a^2} \right\rangle_{\rm average}
 \equiv {1 \over 2\pi ^2}\int {d\Omega_n}
          {1 \over (n\cdot k)^2+a^2}
 = {2 \over a }\cdot{1 \over 
 a + \sqrt {k^2+a^2}} \; .
\label{average}
\eeqn
Here, the dynamical quark is considered to move in various
directions inside hadrons, and hence the constituent quark mass would be regarded as
the quark self-energy in the angle-averaged case\cite{SST2}.
Taking a simple form for the quark propagator 
as $S_q^{-1}(p)=i{p \kern -2mm /}- M(p^2)$,
the SD equation for the quark self-energy $M(p^2)$ is obtained
by taking the trace of Eq.(\ref{sde}) in the Landau gauge 
($\alpha_{e}=0$),
%
%  eq.(8)
%
\barr
M(p^2)&=& \int {d^4k \over (2\pi)^4} {\bf Q}^2 
{M(k^2) \over k^2+M^2(k^2)} \cr
&& \times \left[{2 \over {\tilde k}^2+m_B^2}+{1 \over {\tilde k}^2}
+{4 \over a }\cdot{1 \over 
 a + \sqrt {{\tilde k}^2+a^2}} \left( { {m^2_B-a^2}
\over {{\tilde k^2}+m^2_B} } 
+ {a^2 \over {\tilde k^2}}
\right) \right] 
\label{sde2}
\earr
with ${\tilde k}_\mu \equiv p_\mu - k_\mu$. It is noted that the r.h.s. 
of Eq.(\ref{sde2}) is always non-negative.

We solve Eq.(\ref{sde2}) numerically using the Higashijima-Miransky 
approximation \cite{HM}, ${\bf Q}^2 = 4\pi C_{F} \cdot \alpha_{s} {}^{\rm eff} ( {\rm max} 
\{ p^{2},k^{2} \} )$ with $C_{F}={\footnotesize {{N_{c}^{2} -1} \over {2 N_c}}}$ 
for ${\rm SU}(N_{c})$
in order that the SD equation is reduced to the usual 
one of the QCD-like theory \cite{{HM},{Italy1}} in the ultraviolet region.  
As a consequence, the quark self-energy $M(p^2)$ has the asymptotic 
form, which is obtained by a study of the operator product expansion
and the renomalization group\cite{HM}, 
in the ultraviolet limit. 
The running coupling with the {\it hybrid type} behavior is obtained as
%
%  eq.(9)
%
\beqn
\alpha_{s}{}^{\rm eff}(p^{2})={{12 \pi} \over
{(11N_{c}-2N_{f}) \ln [(p^{2}+p_{c}^{2}) / \Lambda_{\rm 
QCD}^{2}]}}\; ;\;\;
p_{c}^{2}=\Lambda^{2}_{\rm QCD}\exp [{{48\pi^{2}(N_{c}+1)} \over
{(11N_{c}-2N_{f})\cdot e^{2}}}]\;\;,
\label{run}
\eeqn
where $p_{c}$ approximately divides the momentum scale into the 
infrared region and the ultraviolet region 
with the QCD scale parameter $\Lambda_{\rm QCD}$ 
fixed at $200{\rm MeV}$.
Here, $p_{c}$ is fixed by $\alpha_{s}{}^{\rm eff}(p^{2}=0)
=\alpha_{s} / ({N_{c}+1})$; $\alpha_{s}=e^{2}/ 4\pi$.
 The magnitude of the quark condensate plays the role of an order 
parameter of spontaneous chiral-symmetry breaking.
We can calculate the quark condensate with the thus obtained quark self-energy 
$M(p^2)$ from the SD equation as
%
%  eq.(10)
%
\beqn
\langle {\bar q}q \rangle^{^{\Lambda}}
= - {N_c \over 4\pi^{2}}
\int_{0}^{\Lambda^{2}} dk^{2}
{{k^{2}M(k^{2})} \over {k^{2}+M^{2}(k^{2})}}\;\;,
\eeqn
where the ultraviolet cutoff $\Lambda$ is introduced to regularize the 
ultraviolet divergence of the integral. 
If the ultraviolet cutoff is taken in the deeply asymptotic region, $\Lambda \gg 
m_{B}$, one can get the quark condensate in the renormalization-group 
invariant form for $N_{c}=N_{f}=3$, 
$\langle {\bar q}q \rangle_{_{\rm RGI}}=\langle {\bar q}q \rangle^{^{\Lambda}}
\cdot \{ \ln (\Lambda^{2}/ \Lambda^{2}_{\rm QCD}) \} ^{-4/9}$.
The ultraviolet cutoff is taken to be large enough
as $\Lambda / \Lambda_{\rm QCD} = 10^{3}$ hereafter.
One finds that QCD-monopole condensation provides a large 
contribution to spontaneous chiral-symmetry 
breaking\cite{{SST},{TSS},{COMO},{SST2}},
because the mass of the dual gauge field, $m_B$, is proportional to the QCD-monopole 
condensate\cite{{kanazawa},{SST}} and the quark condensate increases 
with $m_B$\cite{{SST},{TSS},{COMO},{SST2}}. In other words, the strength of spontaneous 
chiral-symmetry breaking seems originated from QCD-monopole 
condensation by the sufficient strength of the 
string tension as shown in Fig.~1.
The parameter set is taken as $e=5.5$, $m_{B}=0.5{\rm GeV}$ and $a=85{\rm 
MeV}$, where these parameters except for $a$ 
are set by fitting the confinement phenomena 
at zero temperature. In this case, the quark condensate and the pion decay 
constant at zero temperature are well-reproduced as 
$\langle \bar qq \rangle_{_{\rm RGI}} \simeq-(247{\rm MeV})^3$ and $f_\pi 
\simeq 88{\rm MeV}$, respectively\cite{{SST},{SST2}}.

Now we study the chiral symmetry restoration at finite temperature using 
the imaginary-time formalism for the SD equation\cite{B95}. 
The finite-temperature SD equation is obtained by making the 
following replacement in Eq.(\ref{sde2})\cite{Kapusta}:
%
%  eq.(11)
%
{
\setcounter{enumi}{\value{equation}}
\addtocounter{enumi}{1}
\setcounter{equation}{0}
\renewcommand{\theequation}{\theenumi\alph{equation}}
\jot 0.35cm
\barr
&&p_4 \rightarrow \omega_n = (2n+1)\pi T \; ,\\
&&\int {{d^4 k} \over {(2\pi)}^4} \rightarrow T \sum_{n=-\infty}^{\infty}
\int {{{\rm d}{\bf k}} \over {(2\pi)}^3} \; ,\\
&&M(p^2) \rightarrow M_{_{T}}(\omega_n ,{\bf p}) \; .
\label{FMass}
\earr
\setcounter{equation}{\value{enumi}}
}
\noindent
The resulting equation is very hard to solve even 
numerically since the quark self-energy depends not only on the three 
dimensional momentum $\bf p$, 
but also on the Matsubara frequencies $\omega_n$ \cite{Italy2}.
We propose the {\it covariant-like ansatz}\cite{B95}, in which
a replacement is made for
the quark self-energy at $T\neq0$ instead of (\ref{FMass}) as
%
%  eq.(12)
%
\beqn
M(p^{2})  \rightarrow M_{_{T}}({\bf p}^2 + {\omega^2_n}) 
\equiv M_{_T}(\hat p^2)
\eeqn
with ${\hat p}^2 \equiv {\bf p}^2 + {\omega^2_n}$ and $\omega_n=(2n+1)\pi T$
\cite{B95}. 
It is noted that this ansatz guarantees that the finite-temperature SD equation 
in the limit $T \rightarrow 0$ is exactly reduced to the SD equation (\ref{sde2}) at $T=0$.
This fact is also confirmed by numerical calculations as shown in Fig.(2).
The final form of the SD equation at $T\neq0$ is derived as
%
%  eq.(13)
%
\barr
M_{_T}(\hat p^2) 
&=& {T \over {8 \pi^2}} \sum^{\infty}_{m=-\infty}
\int^{\infty}_{\omega^2_m} {d{\hat k}^2} {\int^{1}_{-1}{dz}} {\bf Q}^2 
\sqrt{\hat k^2 - \omega^2_m}
{M_{_T}(\hat k^2) \over {\hat k^2+M^2_{_T}(\hat k^2)}} \nonumber \\
&& \times \left[ { 2 \over { {\tilde k^2_{nm}} + m^2_B }  } 
+ { 1 \over {\tilde k^2_{nm}} }
+ {4 \over a}\cdot{1 \over {a+
\sqrt{{\tilde k^2_{nm}}+a^2}}} 
\left( { {m^2_B-a^2}
\over {{\tilde k^2_{nm}}+m^2_B} } 
+ {a^2 \over {\tilde k^2_{nm}}}
\right) \right] \;\; ,
\label{sdf}
\earr
where $\tilde k^2_{nm}={\hat k}^2+{\hat p}^2-2z
\sqrt{({\hat p}^2-\omega^2_n)({\hat k}^2-\omega^2_m)}
-2\omega_m\omega_n$. 
It is noted that the large number of the Matsubara frequencies
have negligible contribution at high temperature in Eq.(\ref{sdf})
since $\tilde k^2_{nm} \sim (\omega_m - \omega_n)^{2}$ at $T \gg |{\bf 
p}|, |{\bf k}|$.
In other words, it is important for the small number of the Matsubara frequencies 
to generate the non-trivial solution of the quark 
self-energy $M_{_{T}}(\hat p^2)$ in r.h.s. of Eq.(\ref{sdf}).
We then solve the SD equation by setting $\omega_{n}=0$ in the 
r.h.s. of Eq.(\ref{sdf}).
In addition, it is necessary to truncate the infinite sum of $m$
in order to solve it numerically. 
We check the numerical convergence by this truncation for the 
$m$ sum, which is quite fast at high temperatures.
In the effective running coupling eq.(\ref{run}), there is only a slight modification 
as ${\bf Q}^2=4\pi C_{F}\cdot \alpha_{s}^{\rm eff}
(\max \{ \hat p^{2},\hat k^{2} \})$ due to the covariant-like 
ansatz.
We show in Fig.~2, the quark self-energy $M_{_T}({\hat p}^2)$ as a 
function of ${\hat p}^2$ at finite temperature.
No nontrivial solution is found in the high temperature region, 
$T\stackrel{>}{\scriptstyle \sim}110{\rm MeV}$.  
In other words, the chiral symmetry is restored at high temperature.

The quark condensate is easily calculated using the quark self-energy
$M_{_T}({\hat k}^2)$ as 
%
%  eq.(14)
%
\jot 0.5cm
\barr
\langle {\bar q}q \rangle^{^{\Lambda}}_{_T} 
&=& - 4 N_c \cdot T
\sum_{n=-n_{\rm max}}^{n_{\rm max}}
\int^{\Lambda} {{{\rm d}{\bf k}} \over {(2\pi)}^3}
{M_{_{T}}(\omega_n ,{\bf k}) \over 
{{\bf k}^{2}+\omega_{n}^{2}+M_{_{T}}^{2}(\omega_n ,{\bf k})} } \nonumber \\
&\simeq& - {2N_c \over \pi^2}\cdot T 
\sum_{n=0}^{n_{\rm max}}
\int^{\Lambda^2}_{\omega_n^2} {d{\hat k}^2} \sqrt{\hat k^2 - \omega^2_n}
{M_{_T}(\hat k^2) \over {\hat k^2+M^2_{_T}(\hat k^2)}} 
\earr
with $n_{\rm max} \equiv [ {\Lambda \over {2\pi T}} - {1 \over 2} ]$. 
The quark condensate, $\langle {\bar q}q \rangle
_{_T}^{^{\Lambda}}$, shown in Fig.3, decreases gradually with temperature in the 
low temperature region ($\stackrel{<}{\scriptstyle \sim}0.5T_{_{C}}$)
and vanishes suddenly near the critical temperature $T_{_C}$.
We mention that the behavior of the quark condensate in the low 
temperature region is similar to that of the calculations on the 
chiral perturbation theory, namely $\langle {\bar q}q \rangle_{_{T}} / 
\langle {\bar q}q \rangle_{_{T=0}} = 1 - T^{2}/ {8f^{2}_{\pi}} 
+O(T^{4})$\cite{GL}. 
In addition, these results are slightly different from the lattice QCD 
simulation\cite{Karsch}, which shows that the temperature 
dependence of the chiral condensate is small until 
near the critical temperature.
In these calculations, we do not take any temperature 
dependence in the parameters used
for simplicity, although we should take it into account
as more information on high temperature QCD becomes available.

Finally, we examine the correlation between the critical temperature $T_{_C}$ of the 
chiral symmetry restoration and the confinement quantity such as the 
string tension $\sigma$. As shown in Fig.~4, there exists a strong 
correlation between them.  
From the physical point of view, higher $T_{_C}$ would be necessary to restore the 
chiral symmetry in the system where the linear-confining potential is much 
stronger, because there exists the close 
relation\cite{{SST},{SST2},{Miya},{WO}} 
between color confinement and
dynamical chiral-symmetry breaking through QCD-monopole 
condensation as shown in Fig.~1. 
The critical temperature is estimated 
as  $100{\rm MeV} \stackrel{<}{\scriptstyle \sim}T_{_C} 
\stackrel{<}{\scriptstyle \sim} 110{\rm MeV}$, when one takes the 
standard value of the string tension ($\sqrt{\sigma} \simeq 0.44 {\rm 
GeV}$). 
We remark here that the critical temperature would increase if we take 
smaller cutoff parameter $a$ at finite temperature, which is expected 
as due to the increase of the size of hadrons because of smaller 
confining force\cite{Ichie}.
Hence, it would be very interesting to couple the calculations of 
QCD-monopole condensation and chiral symmetry breaking at finite 
temperature.

In summary, we have studied spontaneous chiral-symmetry breaking and 
its restoration at finite temperature using the SD equation in the DGL theory, 
which is known to 
provide both quark confinement and dynamical chiral-symmetry breaking. 
We have solved the finite-temperature SD equation numerically with the 
covariant-like ansatz.
The chiral symmetry is restored at $T_{_C}\sim100{\rm MeV}$.
We have found the strong correlation between the critical temperature $T_{_C}$ 
and the string tension $\sigma$.

We acknowledge fruitful discussions with O.~Miyamura, S.~Umisedo and 
H.~Ichie on the relation between spontaneous chiral-symmetry breaking
and monopole condensation.
%%%%%%%%%%%%%%%%%%  Reference  %%%%%%%%%%%%%%%%%%%%%%%%%%%%%%

%%%%%%%%%%%%%%%%%%  Figure Caption  %%%%%%%%%%%%%%%%%%%%%%%%%
\newpage

\centerline{\large FIGURE CAPTIONS}

\vspace{0.5cm}

\begin{description}

\item[FIG.1.]
\begin{minipage}[t]{13cm}
\baselineskip=20pt
The correlation between the quark condensate and the mass
of the dual gauge field and also the string tension. 
The other parameters are fixed as $e=5.5$ and $a=85{\rm MeV}$.
The unit is taken by  
$\Lambda_{\rm QCD}\simeq 200{\rm MeV}$.
\end{minipage}

\item[FIG.2.]
\begin{minipage}[t]{13cm}
\baselineskip=20pt
The quark self-energy $M_{_T}({\hat p}^{2})$
as a function 
of ${\hat p}^2$ at $T=0,~60$ and $100{\rm MeV}$
with $e=5.5$, $m_B=0.5{\rm GeV}$ and $a=85{\rm MeV}$.
\end{minipage}

\item[FIG.3.]
\begin{minipage}[t]{13cm}
\baselineskip=20pt
The ratio of 
the quark condensate at finite temperature
to that of zero temperature
as a function of temperature normalized to the critical temperature $T_{_C}$.
The same parameters are used as in Fig.2.
$T_{_C}$ is found at $T_{_C}\simeq 110{\rm MeV}$ in this case.
\end{minipage}

\item[FIG.4.]
\begin{minipage}[t]{13cm}
\baselineskip=20pt
The critical temperature $T_{_C}$ as a function of the square
root of the string tension $\sigma$. 
$T_{_C}$ rises almost linearly with $\sqrt{\sigma}$.
\end{minipage}

\end{description}

\end{document}